\documentclass[12pt,a4paper]{article}
\usepackage{amsmath,amssymb,amsthm,epsfig}
\usepackage{amsfonts}
\usepackage{graphicx}  % needed for figures
\usepackage{dcolumn}   % needed for some tables
\usepackage{bm}        % for math
\usepackage{amssymb}   % for math
\usepackage{amsmath}
\usepackage{epsfig}
\usepackage{hyperref}
\usepackage{cleveref}
\usepackage{datetime}
\newcommand{\N}{\mathbb{N}}
\newcommand{\C}{\mathbb{C}}
\newcommand{\Z}{\mathbb{Z}}
\newcommand{\R}{\mathbb{R}}

\newcommand{\T}{\mathbb{T}}

\newcommand{\cD}{{\cal D}}

\newcommand{\cG}{{\cal G}}
\newcommand{\cH}{{\cal H}}  
\newcommand{\cK}{{\cal K}}

\newcommand{\cM}{{\cal M}}

\newcommand{\cR}{{\cal R}}

\newcommand{\cT}{{\cal T}}

\newcommand{\bear}{\begin{eqnarray}}
\newcommand{\eear}{\end{eqnarray}}

\def\be#1\ee{\begin{equation}#1\end{equation}}
\def\bea#1\eea{\begin{align}#1\end{align}}

\newcommand{\ignore}[1]{}

\newcommand{\Ann}{\mathrm{Ann}\,}
\newcommand{\mspan}{\mathrm{span}\,}
\newcommand{\codim}{\mathrm{codim}\,}
\newcommand{\esp}[1]{\langle #1 \rangle}

\usepackage{marginnote}

\title{
Schwinger-Dyson equations and line integrals
}

\author{
Lorenzo Luis Salcedo$^a$\footnote{email:salcedo@ugr.es}\; 
and\;
Erhard Seiler$^b$\footnote{email:ehs@mpp.mpg.de}\;\;
\mbox{} \\
\mbox{} \\
$^a${\em\normalsize Departamento de F\'{\i}sica At\'omica, Molecular y 
Nuclear}\\ 
{\em\normalsize and  Instituto Carlos I de F\'{\i}sica Te\'orica y 
Computacional}\\  
{\em\normalsize Universidad de Granada}\\
{\em\normalsize Granada, Spain}
\\ 
$^b${\em\normalsize Max-Planck-Institut f\"ur Physik
(Werner-Heisenberg-Institut)} \\
{\em\normalsize Munich, Germany}
} 

\date{\today}

\begin{document}

\maketitle

\begin{abstract} \normalsize \noindent %\hglue-2mm 
The Complex Langevin 
(CL) method sometimes shows convergence to the wrong limit, even though 
the Schwinger-Dyson Equations (SDE) are fulfilled. We analyze this problem 
in a more general context for the case of one complex variable. We prove a 
theorem that shows that under rather general conditions not only the 
equilibrium measure of CL but any linear functional satisfying the SDE on 
a space of test functions is given by a linear combination of integrals 
along paths connecting the zeroes of the underlying measure and 
noncontractible closed paths. This proves rigorously a conjecture stated 
long ago by 
one us (L.~L.~S.) and explains a fact observed in nonergodic cases of CL.

\end{abstract}

\hglue1cm Keywords: Sign problem, Complex Langevin 

\tableofcontents

\newpage
\section{Introduction}

The notorious sign problem arises in many contexts, for instance in the 
functional integral of real time Quantum Mechanics or in imaginary time 
Quantum Field Theory at nonzero chemical potential. The problem comes from 
the fact that even in finite dimensional approximations we are dealing 
with a complex measure whose sign, or rather phase is fluctuating wildly.

A popular method to attack this problem is the Complex Langevin Method 
\cite{Parisi:1984cs,Klauder:1983nn}. It is well known that the method, 
even when it converges, i.~e. defines a stationary probability measure on 
the (complexified) configuration space, may not reproduce the desired 
integrals based on a complex measure.

In this paper we try to give an answer to the question: what does such an 
`incorrect' stationary measure represent then? It is known that the 
expectation values under stationary measures obtained from a Complex 
Langevin process obey certain equations closely related to (but a little 
weaker than) the so-called Schwinger-Dyson Equations characterizing the 
original complex measure.

In order to find an answer we look at a slightly more general problem: 
how can we characterize linear functionals on the space of `observables' 
(functions on configuration space) that satisfy the Schwinger-Dyson 
Equations? It has been conjectured long ago by one of us 
\cite{Salcedo:1993tj} that those functionals are always linear 
combinations of line integrals, where the lines are connecting zeroes of 
the original complex density or are noncontractible closed loops.

In this paper we rigorously prove this conjecture for the one-dimensional 
case under 
some rather general conditions.

%\newpage
\section{Motivation}

\newcommand{\oA}{f}

The purpose of this section is to motivate the theorem of Sec. \ref{sec:3}, so
it is of descriptive character. Precise definitions are postponed to
Secs. \ref{sec:3} and \ref{sec:4}. A one-dimensional setting is assumed
throughout. For definiteness we discuss the nonperiodic setting although
everything can be repeated for the periodic case.

Let $\rho(x)$ be a complex distribution (to be referred to as density) defined
on the real line such that $\int dx \rho(x)=1$.  Given a test function
$\oA(x)$, to extract ``expectation values'' of the type
\be
\esp{\oA} = \int_\R dx \, \rho(x) \oA(x)
\label{eq:esp}
\ee
by means of a Monte Carlo approach constitutes the well-known sign (or in the
present case phase) problem, since plain importance sampling does not apply to
a complex weight $\rho$. 

The Schwinger-Dyson equations (SDE) characterize the density $\rho$ via 
the simple identity between expectation values\footnote{The equations 
refer to $\esp{.}$, not $\oA$. The SDE is the statement that the linear 
form 
\ignore{
$\esp{.}$  defined through $\esp{\oA}\equiv \esp{(\partial_z + v)\oA}$
}
defined through $\oA \mapsto \esp{(\partial_z + v)\oA}$
vanishes identically for some suitably chosen domain of test functions.}
\be
\esp{\oA v}+\esp{f'}=0\,,
\label{eq:sde}
\ee
where
\be
v(x) = \frac{\rho'(x)}{\rho(x)}\,.
\ee
(\ref{eq:sde}) follows from integration by parts under some weak decay 
assumptions.

One way to attack the sign problem when $\rho$ and $\oA$ admit a holomorphic
extension to $\C$ (possibly with isolated singularities or branching points
for $\rho$) is the complex Langevin (CL) approach
\cite{Parisi:1984cs,Klauder:1983nn}. In its simplest version an ensemble of
walkers move on $\C$ with $v(x+iy)$ as drift and with real Gaussian noise
normalized such that the probability density of walkers, $P(x,y;t)$, follows a
Fokker-Planck equation
\be
\frac{\partial P(x,y;t)}{\partial t} = \frac{\partial^2 P}{\partial x^2} 
-
\frac{\partial (v_x P)}{\partial x} - \frac{\partial (v_y P)}{\partial y} ,
\label{eq:fp}
\ee
where $v=v_x+iv_y$ and $t$ is the evolution time parameter.

When the stochastic process reaches a stationary state, $P(x,y)\equiv
P(x,y;\infty)$ defines a linear form $T_P$ on a domain of sufficiently
well-behaved holomorphic test functions as
\be
(T_P,\oA) = \int_\C d^2z\,P(x,y) \oA(z)
.
\ee
Multiplying the {\em stationary} Fokker-Planck equation by $\oA(z)$ and 
using the
Cauchy-Riemann equations for $\oA$, it is straightforward to establish the 
relation
\be
0 = P (\partial_z + v)\partial_z \oA + \partial_x\mathcal{A}_x 
+ \partial_y\mathcal{A}_y.
\ee
The explicit form of $\mathcal{A}_x$ and $\mathcal{A}_y$ is also easily
obtained but not needed here. Therefore, when these boundary terms do not 
contribute upon integration on $\C$, the linear form $T_P$ fulfills the
equations 
\be
0 = (T_P,(\partial_z + v)\partial_z \oA)\,.
\label{eq:cc}
\ee
(\ref{eq:cc}) also follows from using Ito calculus on $\oA(z(t))$, 
averaging over the noise and sending $t\to\infty$ (see for instance 
\cite{Aarts:2009uq}). They characterize the equilibrium behavior, i.~e. 
that expectation values do not change under the CL process, so we call them 
``convergence conditions'' (CC) (cf. \cite{Aarts:2011ax}).

The boundary terms may refer to $\infty$ as an isolated point on the extended
complex plane and also to isolated singularities on $\C$.  In applications the
boundary terms are not guaranteed to drop \cite{Aarts:2011ax,Salcedo:2016kyy},
but they do in some cases such as $\rho(z)$ of the type Gaussian times
polynomial, and others \cite{Aarts:2013uza,Aarts:2017vrv}.

Clearly the linear form $T_\rho$ defined by 
\be 
(T_\rho ,\oA) = \esp{\oA} 
\ee 
not only complies with the the SDE, but the CC as well, for sufficiently 
convergent $\rho$ and $\oA$: 
\be 
\esp{(\partial_z + v)\partial_z \oA } 
%= \int_\R dx \, \rho(x) (\partial_x + v(x))\partial_x \oA(x)
= \int_\R dx \, \rho (\oA^{\prime\prime} + \frac{\rho'}{\rho} \oA') 
= \int_\R dx \, (\rho \oA')' = \rho \oA' \Big|_{-\infty}^{+\infty}\,. 
\ee 
Obviously the SDE imply the CC, provided the test function space is closed 
under taking derivatives. The converse is true if the test function space 
contains with each function also its primitive, for instance if it 
consists of the polynomials; an informal argument for this was given in 
\cite{Aarts:2011ax}. For the periodic case this argument does not hold, 
since the constant function is not the derivative of a periodic function, 
so the space of linear functionals satisfying the CC has one dimension 
more than the one satisfying the SDE. In this paper we will, however, 
always assume that the SDE hold.

The idea of the CL approach is that, provided the boundary terms can be
dropped and the solution of the CC is unique, $T_P$ must coincide with
$T_\rho$ (for sufficient conditions guaranteeing this see
\cite{Salcedo:2015jxd}).  Hence, the expectation values $\esp{\oA}$ can be
computed as averages of $\oA(z)$ weighted with $P(x,y)$, thereby solving the
sign problem. For a more detailed justification of CL, involving
finite evolution times and possible boundary terms arising there, see
\cite{Aarts:2011ax}.

As follows from (\ref{eq:esp}), the form $T_\rho$ relies upon integration
along the real axis of $\rho \oA$.  There are well-known instances in CL in
which the integral does not use the whole real range. For instance
\be
\rho(x) \propto (x-a) e^{-x^2/2}
\ee
for real $a$. If the stochastic process starts at $x_0\in\R$ it remains real
and the segregation theorem applies \cite{nagasawa}, one stationary 
solution $P_+(z) \propto \theta(x-a)\rho(x)\delta(y)$ is obtained if 
$x_0>a$ while another $P_-(z) \propto \theta(a-x)\rho(x)\delta(y)$ obtains 
if $x_0<a$. Therefore the stochastic process is not ergodic 
\cite{Aarts:2017vrv,Seiler:2017wvd}. The two corresponding linear forms 
$T_{P_\pm}$ can be expressed as
\be
(T_{P_\pm},\oA) = \mathcal{N}_\pm \int_{\gamma_\pm} dz\,\rho(z)\oA(z)
\ee
where $\gamma_\pm$ are the paths from $a$ to $\pm\infty$ along the real axis,
and $\mathcal{N}_\pm$ takes care of the normalization. Both linear forms are
solutions of the SDE as is readily verified and $T_\rho$ can be recovered as a
linear combination of them.

Numerical experiments presented in \cite{Salcedo:1993tj} for this $\rho$
indicate that the linear form $T_P$ obtained from CL is a linear combination
of $T_{P_\pm}$ when $a$ lies outside the real axis,\footnote{In this case the
  paths $\gamma_\pm$ go from $a$ to $\pm\infty$ parallel to the real axis.}
even though the stochastic process is ergodic in this case.

%The distribution 
%obtained in the numerical experiment turns out to be is localized not too 
%far from the two paths $\gamma_\pm$ in the support of $P(x,y)$.

Coming back to a generic $\rho$, one can generalize the integral over $\R$
present in $\esp{\oA}$ to other continuous paths $\gamma$ on $\C$ and define
the corresponding linear form as
\be
(T_\gamma ,\oA) = \int_\gamma dz\, \rho(z) \oA(z)
.
\ee
Due to Cauchy's theorem, $T_\gamma$ does not depend on deformations of the
path, provided no singularities are crossed.\footnote{Here we are assuming
  that when $\gamma$ is an open path with finite endpoints, these are not
  affected by the deformation. For endpoints at infinity a sufficient
  condition is to restrict the deformation to a bounded region of $\C$. In
  many cases this is unnecessarily restrictive.} Regarding the SDE, one has
\be
(T_\gamma ,(\partial _z +v)\oA) = 
\int_\gamma dz (\rho \oA)' = \rho \oA\Big|_{\partial\gamma} .
\ee
Hence, there are two cases in which the rhs vanishes and $T_\gamma$ fulfills
the SDE. One is when $\gamma$ connects two zeroes of $\rho$, and such zeroes
are not overruled by the factor $\oA$. This is always true for finite zeroes
and entire holomorphic $\oA$. For a zero at $\infty$, this condition puts
restrictions on the space of test functions. The other case is when $\gamma$
is a closed path; the form $T_\gamma$ is non null if (non removable)
singularities of $\rho$ are enclosed by the path. Denoting $\Gamma$ the set of
paths of the two types just noted, it is clear that (finite) linear
combinations \be T = \sum_{\gamma\in\Gamma} a_\gamma T_\gamma
\label{eq:lc}
\ee
also satisfy the SDE.

Based partially on numerical experiments with CL simulations and partially on
the naturalness of the linear forms $T_\gamma$ in the present context of
holomorphic weights and test functions, it was conjectured in
\cite{Salcedo:1993tj} that the space $\cT$ of linear forms defined in
(\ref{eq:lc}) actually saturates the set of solutions of the SDE for a given
complex density $\rho$.

\newcommand{\minus}{\phantom{-}}
\newcommand{\ip}{\phantom{i\,}}
\newcommand{\iip}{\hspace{5.5pt}}
\begin{table}[t]
\small
\centering
\begin{tabular}{cllll}
%\hline

%\toprule
$\oA(x)$ & ~~~~~~~CL & $a_+ \hat{T}_+ + a_- \hat{T}_-$ 
& ~~~~~~~~~~~$\hat{T}_\pm$  & ~~~~~~$T_\rho$ 
\\
\hline
$x$
& $ \minus i\, 0.5244 (2) $ 
& $ \minus i\, 0.5247 $
& $ \pm 0.7521 + i\, 0.5613 $
& $ \minus i\, 0.9091 $
\\
$x^2$
& $ \minus \ip 0.4129 (9) $
& $ \minus \ip 0.4122 $
& $ \minus 0.3763 \pm i\, 0.7521 $
& $ \minus \ip 0.0284 $
\\
$x^3$
& $ \minus i\, 0.7562 (9)$
& $ \minus i\, 0.7563 $
& $ \pm 0.1880 + i\, 0.7653 $
& $ \minus i\, 0.8523 $
\\
$x^4$
& $ \minus \ip 0.2147 (20) $
& $ \minus \ip 0.2161 $
& $ \minus 0.1733 \pm i\, 0.8931 $
& $ \iip - 0.2397 $
\\
$e^{-ix}$
& $ \minus \ip 1.2100 (6)$
& $ \minus \ip 1.2100 $
& $ \minus 1.2626 \mp i\, 1.0634 $
& $ \minus \ip 1.7544 $
\\
$e^{ix}$
& $ \minus \ip 0.3940 (2)$
& $ \minus \ip 0.3942 $
& $ \minus 0.3754 \pm i\, 0.3808 $
& $ \minus \ip 0.1993 $
\\
$e^{-2ix}$
& $ \minus \ip 0.6109 (21)$
& $ \minus \ip 0.6110 $
& $ \minus 0.7272 \mp i\, 2.3470 $
& $ \minus \ip 1.8126 $
\\
$e^{2ix}$ 
& $ \iip - 0.0064 (3)$
& $ \iip - 0.0063 $
& $ - 0.0186 \pm i\, 0.2491 $
& $  \iip -0.1338 $
\\
\hline
\end{tabular}
\caption{For $\rho(x)=\mathcal{N} (x-i)^2 e^{-\beta x^2}$ with $\beta = 1.6$,
  and several test functions $f$. The columns ``CL'', ``$\hat{T}_\pm$'',
  ``$T_\rho$'' correspond to CL numerical results, $(\hat{T}_\pm,f)$, and
  $\esp{f}$, respectively.  ``$a_+\hat{T}_+ + a_-\hat{T}_-$'' gives the linear
  combination of ``$\hat{T}_\pm$'' with coefficients $a_\pm = 0.5 \mp
  i\, 0.0243(8)$ obtained from a best fit to ``CL''.}
\label{tab:1}
\end{table}

To illustrate the idea we can use a density analyzed in \cite{Aarts:2017vrv}
in the light of CL, namely,
\be
\rho(z) = \mathcal{N} (z-i)^2 e^{-\beta x^2},
\qquad
 \beta= 1.6
\,,
\quad
\mathcal{N} = -0.9634
\,.
\label{ex1}
\ee
Let $\gamma_\pm$ be paths starting at $z=i$ and ending at $z=\pm\infty$. These
generate the set $\Gamma$ of paths. One can define the linear forms
\be
(T_\pm, \oA ) =
\int_{\gamma_\pm} dz \, (z-i)^2 e^{-\beta z^2} \oA(z)
=
\int_0^{\pm\infty} dt \, t^2 e^{-\beta (t+i)^2} \oA(t+i)
,
\ee
with normalizations $(T_\pm,1) = \mp 0.4817 - i\, 0.2228 $. These form a basis
of the space $\cT$ of linear forms. Let $\hat{T}_\pm$ denote the corresponding
normalized versions.

In Table \ref{tab:1} ``expectation values'' of the type $(T,f)$ are displayed
for several choices of $T$ and $f$. The column labeled CL corresponds to $T_P$
of the CL numerical stochastic process. The labels of the other columns are
self-explanatory.  Several ``observables'' $f$ are computed, namely, $x^m$ for
$m=1,2,3,4$, and $e^{ikx}$ for $k=\pm 1,\pm 2$. The CL expectation values are
reproduced with
\begin{equation}
T_P =
a_+ \hat{T}_+  + a_- \hat{T}_-
,
\qquad
a_\pm = 0.5 \mp i\,0.0243(8) 
.
\end{equation}
The relation $a_++a_-=1$ is required by normalization. A single independent
parameter allows to reproduce the various observables obtained from CL. This
is in agreement with the conjecture, and incidentally it checks that the CL
solution does fulfill the SDE for this density. Also noteworthy is the
equivalent form
\begin{equation}
T_P = 
 \frac{b}{2}( \hat{T}_+  + \hat{T}_- )
+(1-b)T_\rho
,
\qquad
b = 1.105(3)
\,.
\end{equation}
Using an overcomplete set of paths, namely, $\gamma_\pm$ and $\R$, and their
normalized linear forms, it has been possible to choose the weights real (but
not nonnegative) for this $\rho$. This fact goes beyond the conjecture.

Many more similar numerical experiments have been done with other
$\rho(z)$. In all cases the forms $T_\gamma$ reproduce the CL numerical
calculation.

The $\rho$ considered include
\be
\rho(z) = (z-z_p)^n e^{-\beta z^2}
\qquad \mathrm{Re}(\beta)>0
\label{eq:branch}
\ee
for various $z_p$ and $\beta$ (including $\beta=1+i$, but mostly $\beta>0$),
and $n=1,2$, but also $n=-1,-2$ and $n=1/2$.  

For integer positive $n$ the case $\rho(x)= (x-i)^2 \exp(-x^2/2)$ is
particularly interesting, since the normalization of $\rho$ vanishes in this
case, $(T_\rho,1)=0$, yet $T_P$ (from CL, which is always normalized) is still
a combination of the two $T_{\gamma_\pm}$.

The negative $n$ introduce poles and in this case the set $\Gamma$ contains
noncontractible closed paths. On the other hand, for noninteger rational $n$
the complex drift $v(z)$ is still univalent so the CL stochastic process takes
place on $\C$. $\rho$ itself is not univalent and the paths $\gamma$ should be
considered to wander on the Riemann surface.  Nevertheless, for a $\rho(z)$ of
the type $(z-z_p)^n\sigma(z)$ with $\sigma(z)$ a univalent function, paths on
different branches but with the same projection on $\C$ give the same
integrals up to a phase, which can be absorbed in the coefficients
$a_\gamma$. Therefore the number of independent paths is still finite, namely,
$2$ for $\rho(z)$ in (\ref{eq:branch}) with $n$ any positive rational number.

Actually, there is no evidence against the conjecture even for more 
general Riemann surfaces, however when $\Gamma$ is not finitely generated 
($\rho$ having an infinite number of zeroes or singularities, or the 
Riemann surface having an infinite number of sheets) the conjecture is 
rendered non predictive, as there would be an infinite number of 
coefficients to be adjusted to reproduce any given linear form $T$ 
fulfilling the SDE. Also in the case of not finitely generated $\Gamma$, 
one would have to enter into functional analysis (topological vector 
spaces), something we want to avoid here. 

Another case analyzed is
\be
\rho(z) = (z-z_1)(z-z_2) e^{-\beta z^2}
,\qquad \beta >0,\qquad z_1\neq z_2\,,
\ee
with two finite zeroes, besides the zeroes at $z=\pm\infty$. Therefore
$\Gamma$ is generated by three independent paths and $\dim\cT=3$. And also
\be
\rho(z) = \exp\left( -\frac{z^2}{2} - \frac{n}{z-z_s} \right),
\ee
with various $z_s$ and $n=1,2$. This $\rho$ has a finite zero at $z_s$ which
is an essential singularity. Hence there are two independent paths connecting
zeroes, plus a closed path around the essential singularity; again
$\dim\cT=3$.

An example of a periodic $\rho(x)$ is provided by the density analyzed in
\cite{Aarts:2017vrv},
\be\begin{split}
\rho(z) &= (1+\kappa\cos(z-i\mu))^{n_p} \exp(\beta\cos(z)),
\\
& \kappa =2,\quad \beta=0.3,\quad \mu=1,\quad n_p=2
\,.
\end{split}\ee
Due to the periodicity of density and test functions, the manifold is
effectively a cylinder, $[0,2\pi]\times \R$. The CL process has two ergodic
components for this $\rho$. There are no (finite) isolated singularities
(hence the only closed paths in $\Gamma$ are the ones winding around the
cylinder) and two finite zeroes at $z=\pm 2\pi/3+i\mu$, which on the cylinder
can be connected in two inequivalent ways (hence $\dim\cT=2$), and can be
combined to form a winding path. These correspond to the two ergodic
components in the present case.

Note however, that in general $\dim \cT>1$ may coexist with an ergodic CL 
process as the example (\ref{ex1}) shows; another example is 
$\rho=\exp(ix^3)$ studied by Guralnik and Pehlevan \cite{Pehlevan:2007eq}.

%\newpage
\section{A general theorem}
\label{sec:3}

\subsection{Nonperiodic case}
\normalsize

We first need a number of definitions.

Let $G_e \subset\C$ be a domain, i.e. a nonempty open connected subset of 
$\C$. We require that the fundamental group $\Pi_1(G_e)$ of $G_e$ is 
finitely generated. This implies that $G_e=\bar\C\setminus 
\cup_{i=1}^{n_r}R_i$
~($n_r<\infty)$, where the $R_i$ are connected and simply connected closed
disjoint subsets of $\bar\C$ and $\bar \C=\C\cup\{\infty \}$ is the Riemann
sphere.

{\em Remark 1:} $G_e \subset\C$ implies that $\infty$ is always an element of
one of the $R_i$. The $R_i$ can be single points.

{\em Definition 1 (set of densities):} $\cR(G_e)$ is the set of functions 
(``densities'') $\rho$ which are meromorphic on $G_e(\rho)$, with a set 
$P(\rho)$ of poles. We denote by $G(\rho)=G_e(\rho)\setminus P(\rho)$ 
the domain of holomorphy of $\rho$.

$\rho$ will in general have zeroes in $G(\rho)$; we denote the set of zeroes
by $N_0(\rho) $; The sets $P(\rho)$ and $N_0(\rho)$ are required to be finite.

{\em Remark 2:} $P$ being finite, the fundamental group $\Pi_1(G)$ of $G$ 
is also finitely generated.

{\em Definition 2 (boundary zeroes):} Let $\partial G_e$ be the boundary of
$G_e$ in $\bar\C$.  We say that $\rho$ has a boundary zero at $b \in \partial
G_e$ iff there is a smooth path $\gamma$ in $G$ such that \be \lim_{z\to
  b|z\in\gamma} \rho(z)=0\,.  \ee We denote the set of boundary zeroes by
$N_e(\rho)$ and the total set of generalized zeroes by $N(\rho)\equiv
N_0(\rho)\cup N_e(\rho)$.

{\em Remark 3:} $\infty$ can be at the boundary of $G_e$, hence it can be an
element of $N_e$. Essential singularities of $\rho$ are instances of boundary
zeroes, $R_i$ being a single point in those cases.
%\marginnote{\bf Is this always true?}

Next we define a set of paths as follows:

{\em Definition 3 (maximal set of paths):} Given $\rho\in\cR(G_e)$, the 
set $\Gamma_m(\rho)$ consists \ignore{of the set} of all oriented open and 
all closed noncontractible paths $\gamma$ in $G$ \ignore{$G\cap \C$} such that
\begin{itemize}
\item 
the closed noncontractible paths are oriented rectifiable curves, 
differentiably parameterized by arc length: $[s_-,s_+]\ni s\mapsto z_\gamma(s)$ 
with $z_\gamma(s_-)=z_\gamma(s_+)$,

\item
the arcs $\gamma$ are oriented curves, differentiably parameterized by arc
length: $(s_-,s_+)\ni s\mapsto z_\gamma(s)$ where now $s_\pm$ may be $\pm
\infty$ and the endpoints are elements $N(\rho)$, that is, the limits
$z_\pm(\gamma) \equiv \lim_{s\to s_\pm}(z_\gamma(s))$ exist in $\bar\C$ and
$z_\pm(\gamma) \in N(\rho)$.
\end{itemize}

{\em Definition 4 (weak and strong decay):} Let $f$ be holomorphic in 
$G_e$ and $z_0\in N(\rho)$ be a generalized zero of $\rho\in \cR(G_e)$; 
let $\gamma\in\Gamma_m(\rho)$ be such that $\lim_{s\to s_+} 
z_\gamma(s)=z_0$. $f$ decays weakly at $z_0$ iff
\be
\lim_{s\to s_+} f(z_\gamma(s))\rho(z_\gamma(s))=0   
\label{weak}
\ee
and it decays strongly iff
\be
\lim_{s\to s_+} |s|^\alpha f(z_\gamma(s))\rho(z_\gamma(s))=0\quad 
\forall \alpha\in\R\,.
\label{strong}
\ee
Analogous definitions apply when $z_0=\lim_{s\to s_-} z_\gamma(s)$.

{\em Definition 5 (space of test functions):} Let $\Gamma$ be a subset of 
$\Gamma_m(\rho)$. Then we define $\cD(\Gamma)$ as the set of functions 
holomorphic in $G_e$ which are of weak decay at the finite endpoints of 
$\gamma\in\Gamma$ and of strong decay at the infinite generalized zeroes 
of $\rho$ which are endpoints of a $\gamma\in\Gamma$.

{\em Definition 6 (Schwinger-Dyson operator):} We define the SD operator 
$A$ on $f\in \cD(\Gamma)$ by 
\be
(Af)(z)\equiv  
f'(z)+\frac{\rho'(z)}{\rho(z)}f(z)=\frac{1}{\rho}\partial_z(\rho(z) f(z))\,.
\ee 
Note that $Af$ is in general no longer holomorphic, but meromorphic in 
$G(\rho)$, so as codomain of $A$ we have taken the space of meromorphic 
functions on $G_e$.

{\em Definition 7 (SD space):} We define the `Schwinger-Dyson space' by
\be
\cH\equiv \cD+A\cD\,.
\label{hspace}
\ee

{\em Definition 8 (SDE):}
We say that a linear functional $T$ on $\cH$ satisfies the Schwinger-Dyson 
equations  iff
\be
(T,Af)=0\;\;\forall f\in \cD\,.
\label{sde}
\ee 
{\em Definition 9 (linear functionals given by paths):} For any path 
$\gamma\in\Gamma$ and $f\in \cH(\Gamma)$ we define 
\be
(T_\gamma,f)\equiv \int_{\gamma} dz \rho(z)f(z)\,;
\ee
%For a set $\Gamma$ of paths 
We define $\cT(\Gamma)$ to be the linear span 
of the $T_\gamma, \gamma\in \Gamma$.

{\em Remark 4:} The linear forms in $\cT(\Gamma)$ fulfill the SDE, that is,
they annihilate $A\cD(\Gamma)$.

Now we can formulate

{\bf Theorem 1:} Let $\Gamma=\{\gamma_1,\ldots\gamma_n \}$ be a finite subset
of $\Gamma_m(\rho)$, such that \\
i) the set of open paths in $\Gamma$ forms a connected network,\\ 
ii) each zero in $N_0(\rho)$ is the endpoint of an open path in $\Gamma$, 
and \\ 
iii) the set of closed paths in $\Gamma$ generates the fundamental group 
$\Pi_1(G)$ of $G$. \\ 
Then any linear functional $T$ on $\cH$ satisfying the SDE (\ref{sde}) is 
given by a linear combination of integrals of $f\rho$ over paths in 
$\Gamma$, i.~e. 
\be
(T,f)= \sum_{i=1}^n a_i (T_{\gamma_i},f)
\ee
with some coefficients $a_i\in\C$.

To prove the theorem we need the following

{\bf Lemma 1:} Let $\Gamma$ be as above and $f\in\cH$ such that
\be
(T_\gamma,f)=0 \quad \forall \gamma\in\Gamma \,,
\ee
then $f\in A\cD$.

{\em Proof:} Because $\cH=\cD+A\cD$  and $T_\gamma$ annihilates $A\cD$,
 it is sufficient to consider the case
$f\in\cD$. It has to be shown that $f\rho=(h\rho)'$ for some $h\in\cD$.
Let $N'$ be the set of generalized zeroes connected by some path in
$\Gamma$. By assumption $N_0\subset N'\subset N$.

Let us assume first that the set $N'$ is empty. Let $z_0\in G$ and
\be
F(z)=\int_{z_0}^z dz' f(z')\rho(z')\,,
\ee
so that $h=F/\rho$ is a solution of the differential equation 
$f\rho=(h\rho)'$. The condition that the closed paths in $\Gamma$
generate $\Pi_1(G)$ implies that $F$, and hence $h$, is univalent because
\be
\oint_\gamma f(z)\rho(z) dz=0\,
\ee
for closed paths $\gamma\in\Gamma$ around each $R_i$ and each pole of $\rho$.
In addition, where $\rho$ has a pole of order $m$ $F$
has at most a pole of order $m-1$, hence $h$ is holomorphic in $G_e$.

When $N'$ is not empty we define $F$ as above with $z_0\in N'$. $F$ vanishes
at $z_0$ and also at all other generalized zeroes in $N'$ since they are
connected directly or indirectly to $z_0$ by open paths $\gamma\in\Gamma$
fulfilling
\be
\int_\gamma f(z)\rho(z) dz=0\,.
\ee
It follows that $h=F/\rho$ has weak decay at the finite zeroes of $N^\prime$.
$h$ also has strong decay at the infinite (generalized) zeroes of $N^\prime$,
because the integral of a strongly decaying function approaches its limit
faster than any power. Finally, $h$ is holomorphic at $N_0(\rho)$ because
where $\rho$ has a zero of order $m$ $F$ has a zero of order at least
$m+1$. Therefore $h\in\cD$. The solution $h$ is unique iff $N'$ is nonempty.
\hfill $\Box$

{\em Proof of Theorem 1:} 
We need some preparation:

Let $\cK$ be the subspace of $\cH$ consisting of the $f\in\cK$ satisfying
\be
(T_\gamma, f)=0 \quad \forall \gamma\in\Gamma\,.
\ee
$\cK$ is determined by $n$ not necessarily linearly independent 
linear conditions. We  pick a subset $\Gamma_0$ of $\Gamma$ such that
the set $\{T_\gamma|\gamma\in\Gamma_0\}$ is a basis of $\cT(\Gamma)$.  
$|\Gamma_0|=n_0\le n$ is then the dimension of the quotient 
space $\cH/\cK$. It follows that $\cH$ can be written as a direct sum 
\be
\cH=\cK\oplus\cG
\ee
with $\dim\cG=\dim \cH/\cK=n_0$. ($\cG$ is of course not unique). So any 
$f\in\cH$ can written as 
\be
f = f_\cK + f_\cG
\ee
with $f_\cK\in\cK$ and  $f_\cG\in\cG$.
Once $\cG$ has been chosen, the dual space $\cH^*$ of $\cH$ can be split 
accordingly: 
\be
\cH^*=\cK^*\oplus\cG^*
\ee
and any $T\in\cH^*$ can be written as 
\be
T = T_\cK + T_\cG
\ee
with $T_\cK\in\cK^*$ and  $T_\cG\in\cG^*$, such that
\be 
(T_\cG,f_\cK)=0=(T_\cK,f_\cG)\,.
\ee
Note that we understand here $\cK^*$ and $\cG^*$ as spaces of linear 
functionals acting on $\cH$. $\cG^*$ is then the subspace of $\cH^*$ 
annihilating $\cK$. We now state

{\bf Lemma 2:} $\cG^*$ is spanned by the set $\{T_\gamma|\gamma\in 
\Gamma_0\}$, so $\cG^*=\cT(\Gamma)$.

{\em Proof:} The set $\{T_\gamma|\gamma\in\Gamma_0\}$ annihilates $\cK$ by 
definition, so it is a subset of $\cG^*$; since it consists of $n_0$ 
linearly independent elements it spans all of $\cG^*$. \hfill $\Box$

To complete the proof of the theorem we note that, by Lemma 1, $\cK\subset
A\cD$.  Obviously also $A\cD \subset \cK$, 
so we have 
\be
\cK=A\cD\,.
\ee
Now let $T$ obey the SDE, i.e. Eq.(\ref{sde}) holds. But then also 
\be
(T,f)=0\quad \forall f\in\cK\,.
\ee 
i.e. $T\in\cG^*$. So Theorem 1 follows by using Lemma 2. 
\hfill $\Box$

{\em Remarks:}\\ {\em 5.} We can enlarge $\Gamma$ considerably by allowing homotopic
deformations in $G(\rho)$ of the paths $\gamma$, provided they leave $\cD$ and
$T_\gamma$ invariant. This is guaranteed if the deformation takes place
outside an open neighborhood (in $\bar\C$) of the endpoints when they are
boundary zeroes of $\rho$ (see Remark 8 below).
\\{\em 6}. We may also enlarge $\cD(\rho)$ by completing this space with
respect to some (semi)norms. For instance we can introduce an $L^1$ norm on
$\cD(\rho)$ by \be ||f||_1\equiv \sum_{\gamma\in\Gamma_0} \int_\gamma
|f(z)\rho(z)||dz| \,; \ee which is finite for $f\in\cD$, and define the
completion $\bar\cD$ with respect to $||.||_1$.  Clearly $\cT(\Gamma)$ will be
continuous in that norm. Theorem 1 remains true if we replace $\cH^*$ (the
algebraic dual of $\cH$) by $\cH'$ (the topological dual of $\cH$).
\\{\em 7}. Lemma 1 does not require $\Gamma$ to be finite. Hence Theorem 1
would still hold for infinite sets $\Gamma$, provided $\cT(\rho)$ is
finite-dimensional (required for Lemma 2).
\\{\em 8}. The paths in $\Gamma$, a subset of $\Gamma_m(\rho)$, can be bundled into
equivalence classes through the relation $\gamma_1\equiv\gamma_2$ iff
$T_{\gamma_1}=T_{\gamma_2}$. However, that an open path is
contractible\footnote{Contractible here means that the two endpoints coincide
  and the closed path in $\bar\C$ obtained by adding the endpoint, $z_0$, is
  contractible within $G\cup\{z_0\}$.} does not imply it is null when the
endpoint is a boundary zero of $\rho$.  Likewise, two homotopic open paths
need not be equivalent if any of the endpoints is a boundary zero. The two
paths will be equivalent if they coincide in an open neighborhood (with
respect to $\bar\C$) of their endpoints.

\subsection{Periodic case}

Let $\T$ be the cylinder $[0, 2\pi]\times \R$ with $0$ and $2\pi$ identified.
We can compactify $\T$ to $\bar\T$ by adding two points $+i\infty$ and
$-i\infty$ to obtain a topological sphere. Let $G_e$ be a nonempty open
connected subset of $\T$ such that $\Pi_1(G_e)$ is finitely generated.

{\em Definition 1$\kern 0.1em '$ (set of periodic densities):} $\cR_p(G_e)$ is
the set of functions (``densities'') which are meromorphic on $G_e(\rho)$,
with a finite set $P(\rho)$ of poles and a finite set $N_0(\rho)$ of zeroes.
We denote by $G(\rho)=G_e(\rho)\setminus P (\rho)$ the domain of holomorphy of
$\rho$.

{\em Remark 9:} By definition the densities $\rho\in\cR_p(G_e)$ can be 
viewed as functions on (a subset of) $\C$ satisfying 
\be
\rho(z + 2\pi) = \rho(z) . 
\ee
We have to consider the periodic case separately because of our finiteness 
assumptions made for the nonperiodic case. They would be violated if we 
considered the periodic case simply as a special case of the nonperiodic 
one.

{\em Definition 2$\,'$ (boundary zeroes):} Let $\partial G_e$ be the boundary of
$G_e$ in $\bar\T$.  We say that $\rho$ has a boundary zero at $b \in \partial
G_e$ iff there is a smooth path $\gamma$ in $G$ such that
\be \lim_{z\to
  b|z\in\gamma} \rho(z)=0\,.
\ee
We denote the set of boundary zeroes by $N_e(\rho)$ and the total set of
generalized zeroes by $N(\rho)\equiv N_0(\rho)\cup N_e(\rho)$.

We then have the periodic analogue of Theorem 1:

{\bf Theorem 1$'$:} Let $\Gamma=\{\gamma_1,\ldots\gamma_n \}$ be a finite 
subset of $\Gamma_m(\rho)$, such that 
\\i) the set of open paths in $\Gamma$ forms a connected network, 
\\ii) each zero in $N_0(\rho)$ is the endpoint of an open path in $\Gamma$, 
and
\\iii) the set of closed paths in $\Gamma$ generates the fundamental group of
$G$.  
\\Then any linear functional $T$ on $\cH$ satisfying the SDE (\ref{sde}) 
is given by a linear combination of integrals of $f\rho$ over paths in 
$\Gamma$,

\be
(T,f)= \sum_{i=1}^n a_i (T_{\gamma_i},f) 
\ee
with some coefficients $a_i\in\C$.

{\em Proof:} The proof of Theorem 1 applies with some obvious adjustments.
\hfill $\Box$

%\newpage
\section{Special instances of the theorem}
\label{sec:4}

\subsection{Preliminaries}

In the previous section we have constructed a large class of triples
$(\rho,\Gamma,\cD)$ for which the fulfillment of the SDE (\ref{sde}) implies
that the functional is a combination of pathwise functionals: if $T\in \cH^*$
such that $(T,A\cD)=0$ then $T\in \cT(\Gamma)$. Equivalently
\be
\Ann(A\cD) \subset \cT \,,
\label{theorem}
\ee
where $\Ann$ denotes the annihilator space of a set $S\subset \cH$, i.e., 
\be
\Ann(S) \equiv \{T\in \cH^*\,|\, (T,f)=0 ~\forall f\in S\}
.
\ee
The method to prove this started by showing that $\cK=A\cD$ (Lemma 1) where
$\cK \equiv \cap_{\gamma\in\Gamma}\ker T_\gamma$, and then showing that
$\Ann(\cK) = \cT$ when $\dim\cT<\infty$. In turn, Lemma 1 is based on
$\cD\cap \cK \subset A\cD$, or in words, if $f\in\cD$ fulfills the constraints
$(T_\gamma,f)=0$~$\forall \gamma\in\Gamma$, the equation $f=Ah$ has a solution
$h\in \cD$.

It is natural to seek other triples of density, set of paths and domain of $A$
fulfilling the relation (\ref{theorem}) which can be regarded as a corollary
of the theorem. A way to do this is as follows. Given $\rho\in\cR$ and
$\Gamma$ a finite subset of $\Gamma_m(\rho)$, one can consider triples
$(\rho,\Gamma,\cD_1)$, where now a subspace $\cD_1\subset\cD$ is used as
domain of $A$. (This introduces the corresponding $\cH_1=\cD_1+A\cD_1$,
$\cT_1\subset\cH_1^*$, etc.) The triple will continue to be consistent in the
sense that $\cT_1\subset\Ann(A\cD_1)$ but in general
$\Ann(A\cD_1) \subset \cT_1$ will not hold for such smaller domain. The
impediment comes from Lemma 1: in general, $f$ of $\cD_1$ fulfilling the
constraints $(T_\gamma,f)=0$~$\forall \gamma\in\Gamma$ implies that $f=Ah$ for
some $h\in\cD$ but not necessarily $h\in\cD_1$. In some sense $\cD_1$ is not
complete. One can remedy this by enlarging the domain to $\cD_2$ by simply
adding all the new $h$ so obtained. The new $\cD_2$ still will be not complete
in general, but the process of enlarging the domain can be iterated until
reaching a fixed point $\cD_\infty$. This is the smallest domain with
$\cD_1\subset\cD_\infty\subset\cD$ such that the triple
$(\rho,\Gamma,\cD_\infty)$ is complete, i.e.,
$\Ann(A\cD_\infty) \subset \cT_\infty$ holds.\footnote{It can also be noted
that the previous discussion holds whenever the triple $(\rho,\Gamma,\cD)$ is
complete, possibly with a domain $\cD$ different from that considered in
Sec. \ref{sec:3}.}

The domain $\cD_\infty$ can be $\cD$ itself, but not necessarily. In this 
section
we will show that for large classes of $\rho$ simple domains, like polynomials
in the noncompact case or Fourier modes in the periodic case, yield triples
which are complete.

To analyze such cases we will make use of an alternative proof of the relation
(\ref{theorem}), where $\cD$ refers now to the domain of $A$ considered (in
general a subspace of the large domain given in Definition 5 of
Sec. \ref{sec:3}). The alternative way of proving the result in
(\ref{theorem}) is to show that 1) $\cT \subset \Ann(A\cD)$ and 2) $\dim\cT =
\dim\Ann(A\cD)$. When $\codim A\cD = \dim\Ann(A\cD) < \infty$ 
this immediately implies $\cT = \Ann(A\cD)$ and hence (\ref{theorem}).
Letting
\be
N_\Gamma \equiv \dim \cT,
\qquad
N_{\mathrm{SDE}} \equiv \codim A\cD
,
\ee
the key relation is
\be
N_{\mathrm{SDE}} = N_\Gamma
,
\ee
and in fact showing $N_{\mathrm{SDE}} \le N_\Gamma$ suffices.

\subsection{Densities of rational type}

We make the following definitions:

{\em Definition 10 (Nonperiodic densities of rational type):}
A density $\rho\in\cR$ is called {\em of rational type} iff 
$\rho'(z)/\rho(z)$ is a rational function. 

{\em Definition 10$\,'$ (Periodic densities of rational type):}
A density $\rho\in\cR_p$ is called {\em of rational type} iff
$\rho'(z)/\rho(z)$ is a rational function of $\omega=\exp(iz)$.

For these classes of densities, with rather natural choices of $\cD$ and
$\Gamma$, it is possible to determine explicitly the numbers
$N_{\mathrm{SDE}}$ and $N_\Gamma$ and show that they are equal. We provide the
general, rather elaborate discussion in Appendix \ref{appA} and limit
ourselves here to three examples illustrating how the new strategy (of proof)
works and also to point out the difference with the previous strategy.

{\em Example 1:} This is an extremely simplified case, in the periodic 
setting: The triple $(\rho,\Gamma,\cD)$ is composed of $\rho(z)=e^{-iz}$, 
$\Gamma$ contains just the path $\gamma$ winding once around the cylinder 
$[0,2\pi]\times\R$ (in the positive real direction) and $\cD$, the domain 
of $A$, is the space spanned by the Fourier modes:
\be
 \cD = \mspan\{e^{ikz}, k\in\Z\}\,.
\label{eq:29}
\ee
Applying the SD operator $A$ on a basis of $\cD$,
\be
Ae^{ikz} = i(k-1)e^{ikz}
\qquad k\in\Z \,,
\ee
yields the image of $A$
\be
A\cD = \mspan\{e^{ikz}, k\in\Z\setminus\!\{1\} \, \}\,.
\ee
In this example $\cH=\cD+A\cD$ is just $\cD$, and the dual space $\cH^*$ 
consists of all sequences of complex numbers.  
As it is readily verified $(T_\gamma,f)=0$ for all $f\in A\cD$, hence
$\cT\subset\Ann(A\cD)$. Also the space
\be
\cK = \ker T_\gamma = \{ f\in\cH \, |\, \oint_\gamma dz \rho f  =0 \}
\ee
coincides with $A\cD$ (so Lemma 1 is satisfied). To check that $\Ann(\cK) 
= \cT$ ($\cT$ being finite-dimensional) it is sufficient to verify that their
dimensions coincide, i.e., $\dim\Ann(\cK)=1$ in this case. This is equivalent
to $\dim\cH/\cK=1$. And indeed, one can write the decomposition
\be
\forall k\in\Z \qquad e^{ikz} = (1-\delta_{k,1})e^{ikz} + \delta_{k,1}e^{iz}
\ee
where the first function in the r.h.s. is in $\cK$ and the second one is in a
one-dimensional complementary space of $\cK$ (denoted $\cG$ in the proof of
Theorem 1). From here the theorem $T\in\Ann(A\cD) \Longrightarrow T=\lambda
T_\gamma$ follows:
\be
T,T_\gamma \in \Ann(\cK)
\qquad
(T,e^{ikz})  = (T,e^{iz}) \delta_{k,1} ,
\qquad
(T_\gamma,e^{ikz})  = 2\pi \delta_{k,1}
\ee
hence  $(T,f)=\lambda (T_\gamma,f)$ with $\lambda = (T,e^{iz})/(2\pi)$.

In the alternative proof $\cK$ is not explicitly used; $N_\Gamma=1$ as there
is just one path, and $N_{\mathrm{SDE}}$ is directly computed by looking for
the most general $T\in \cH^*$ fulfilling the SDE:
\be
0 = (T,Ae^{ikz}) = i(k-1)(T,e^{ikz}) \,.
\ee
Clearly, all the components $(T,e^{ikz})$ of $T$ except $k=1$ are determined
by the SDE (namely, they vanish in this case). Therefore $N_{\mathrm{SDE}}=1$, which
coincides with $N_\Gamma$, implying $\Ann(A\cD)=\cT$ (using $\cT\subset\Ann(A\cD)$).
\hfill $\Box$

{\em Example 2:} This is a slightly more complicated case in the the 
noncompact setting. Let
\be
\rho(z) = \frac{(z-a)^3}{(z-b)^2}e^{-z^4-1/z^2}
\qquad
(a\neq b)
\label{eq:sp2}
\ee
with $\cD$ the space of polynomials, 
\be   
 \cD = \mspan\{z^k, k\in\N_0\}\,,
\ee   
The image $A\cD$, as well as $\cH=\cD+A\cD$ is living in the space of 
rational functions.  
There is an essential singularity at $z=0$ and a pole at $z=b$ unless
$b=0$. So the number of closed paths is $N_c=1$ if $b=0$ and $N_c=2$
otherwise. (It is important to note that singularities at infinity do not 
add
any closed paths.)  There are zeroes at $z=0,a,\infty$. But the zero at
infinity is reached in four different (path-inequivalent) ways giving rise 
to
an effective number of four zeroes. Similarly, the zero at the essential
singularity at $z=0$ can be reached in two inequivalent ways, namely, 
$z\to
0^+$ or $z\to 0^-$, so from the point of view of $N_\Gamma$ it counts as 
two
zeroes.\footnote{For instance, the path $z_\gamma(t)= i+e^{it}$, $-\pi/2< 
t <
3\pi/2$, defines a linear form $T_\gamma\not\equiv 0$ on the space of
polynomial test functions, regardless of the values of $a$ and $b$.}
Effectively the number of zeroes is $N_z=7$, unless $a=0$ in which case
$N_z=6$. This produces $N_z-1$ linearly independent paths connecting
them. Hence, $N_\Gamma= N_c+N_z-1$. One last observation is that $N_z-1$ is
the number of open paths because $N_z>0$. When there are no zeroes the 
number of open paths connecting them is itself zero. As for the evaluation of
$N_{\mathrm{SDE}}$ for the density in (\ref{eq:sp2}), it is not simpler than for
the general case. The computation in Appendix \ref{appA.1} shows that
$N_{\mathrm{SDE}}$ indeed coincides with $N_\Gamma$.
\hfill $\Box$.

{\em Example 3.} We consider a less trivial periodic example, with the same
domain as in (\ref{eq:29}): Let 
\be
\rho(z)= e^{\sigma i z}\exp(e^{2iz})
\ee
with $\sigma=\pm 1$. There are two zeroes at $z=-i\infty$ from $e^{2iz}$,
hence $N_\Gamma=2$.

It can be noted that when $\sigma=1$ $\rho$ has an additional zero at
$z=+i\infty$ which however does not contribute to $N_\Gamma$ because it fails
to be a zero common to all functions in $\rho\cD$: $N_\Gamma$ counts the
number of independent paths $\gamma$ for which $T_\gamma$ fulfills the SDE as
linear forms on a given domain $\cD$ of test functions. For paths connecting
zeroes, SDE requires $\rho f$ to vanish at the zeroes for all $f\in\cD$
(the vanishing of $\rho$ is not sufficient). Clearly a factor $f=e^{ikz}$ with
$k\le -1$ cancels the zero of $\rho $ at $z=+i\infty$ for $\sigma=1$ in the
product $\rho f$.

Let $\omega\equiv e^{iz}$.
Application of the SD operator on a generic element of the basis produces
$-iA\omega^n = (n+\sigma)\omega^n + 2 \omega^{n+2}$. If $T$ complies with 
the
SDE, $0=(T,A\omega^n)$, this gives
\be
0 = (n+\sigma) E_n + 2 E_{n+2} \qquad \forall n\in\Z \,,
\ee
where $E_n\equiv(T,\omega^n)$. Clearly, all $E_n$ with even $n$ are 
determined
by $E_0$. On the other hand for $\sigma=1$ $E_n=0$ for all positive odd 
$n$  
and all $E_n$ with odd $n<-1$ are determined by $E_{-1}$, hence there are 
two
arbitrary coefficients and $N_{\mathrm{SDE}}=2$, matching $N_\Gamma$. For
$\sigma=-1$ the situation is similar. \hfill $\Box$

These examples, while being nongeneric, still serve as the paradigms for 
the treatment of the generic cases of rational type, described in 
Appendix \ref{appA}.

%\newpage
\section{Conclusions and outlook}

Motivated by observations in the study of complex Langevin simulations we 
proved under rather general conditions that for one-dimensional systems 
with a complex density $\rho$ any linear functional on the `observables' 
satisfying the SDE is a linear combination of line integrals, where the 
lines either connect zeroes of $\rho$ or are noncontractible closed loops.

For the nonperiodic case, this implies that the stationary expectation 
values obtained by the Complex Langevin process, which satisfy 
superficially weaker conditions (the `convergence conditions' CC) are also 
given by such linear combinations.

For the periodic case we can only prove a weaker statement, because in 
this case the CC do not imply the SDE. The weaker statement is that those 
expectation values are linear combinations of the line integrals as above 
plus one extra linear functional which is not of that form.

The single mode that in the periodic case is not a derivative is the constant
mode $f=1$, hence the SD condition that is not a consequence of CC in the
steady CL solution is $0=(T_P,A1)=(T_P,v)$. As shown in Appendix \ref{appB}
under reasonable conditions flux conservation guarantees the property 
$0=\mathrm{Im}(T_P,v)$, which is close to what would be needed.
The correct relation $(T_P,v)=0$ is checked for the density $\rho(x) =
\exp(imx + \beta \cos(x))$ ($m\in\Z$) analyzed in \cite{Salcedo:2016kyy} in
two cases, $\beta\in\C$ with $m=0$, and $\beta\in\R$ with $m\neq 0$. While in
the former case $(T_P,v)$ vanishes due to parity, the vanishing in the second
case is not trivial.\footnote{For these densities $v(z)=im-\beta\sin(z)$ and
  the integral $\int d^2z\,P(z)v(z)$ is absolutely convergent, as shown in
  \cite{Salcedo:2016kyy}. This is no longer guaranteed for higher Fourier
  modes, $e^{ikz}$ with $|k|>1$.}

It should be remarked that in the study of Complex Langevin we do not know 
of a single case in which such an extra linear functional actually 
appears.  A possible reason is the fact observed in \cite{Aarts:2011ax} 
that a linear functional satisfying the CC together with some bounds in 
terms of a sup norm of the observables on a path imply (via the 
Riesz-Markov theorem) that the functional is indeed given by the 
corresponding line integral. But the issue is far from being settled and 
needs further study.

An important open question concerns the generalization of our results to 
higher dimensions, in particular group manifolds, which form the 
configuration space of lattice gauge theories. We think that this is 
probably possible, but not quite trivial since it will involve the theory 
of analytic functions of several complex variables.

It can be noted that the SDE operator $A=\partial_z+v(z)$ has the form of a
(gauge) covariant derivative. In fact a kind of gauge symmetry is present in
the problem analyzed here; letting $\omega(z)$ be a holomorphic function
without zeroes (hence $\omega^{-1}$ being also holomorphic), the space of test
functions is unchanged under the transformation $\tilde{f} = \omega f$. In
this case, the identity $\int f \rho \,dx = \int (f \omega) (\omega^{-1}\rho)
\, dx$, indicates that $\tilde\rho = \omega^{-1}\rho$ is gauge equivalent to
$\rho$ and this property is preserved by the SDE: $0=\int dx \partial (\rho f)
= \int dx \partial (\tilde\rho \tilde f)$. The upshot is that if the
statement in Eq. (\ref{eq:lc}) must hold for a given $\rho$, it must hold too
for any other complex density $\tilde\rho$ in the same gauge orbit. Such gauge
copies share in common the position of zeroes and singularities. Going a bit
further one can consider the transformation $\tilde{f} \to \rho f$, so that
$\tilde{\rho}=1$ and the SDE operator becomes simply
$\tilde{A}=\partial_z$. The domain of test functions becomes
$\tilde{\cD}=\rho\cD$ ($\cD$ being the original domain) and in this setting
all the information about the gauge orbit is contained in the domain
$\tilde{\cD}$, which is composed of analytic functions with some prescribed
zeroes and singularities. That structure, being common to all functions, can
be regarded as a property of the manifold itself, and the property
$\tilde{A}=\partial_z$ suggests the speculation that the problem analyzed in
this work could be reformulated as one of determining the cohomological
properties of that manifold.

This idea is reinforced (through the presence of the exterior derivative) when
another symmetry, general covariance, is considered. The basic structure
$\esp{f}=\int f \rho \,d^n x$ (in a general $n$-dimensional case) has a
geometric, i.e. coordinate independent, meaning where $f$ is a $0$-form and
$\hat{\rho}\equiv \rho \, d^nx$ is an $n$-form. However the SDE are not
automatically general covariant: for each coordinate $x^i$, $i=1,\ldots,n$,
one can consider an SD operator $A_i$ defined by
$A_if=\rho^{-1}\partial_i(\rho f)$, and indeed $\esp{A_if}=\int
d^nx\,\partial_i(\rho f)$ must vanish, assuming sufficient convergence at the
boundary, but the $A_i$ depend on the coordinate system.
A covariant operator can be written as 
\be
A_\xi f = \rho^{-1} \partial_i (\xi^i \rho f),
\ee
where $\xi^i(x)$ is a vector field, leading to
$\esp{A_\xi f}=0$. The SDE express the invariance of $\int f \rho \,d^n x$
under rearrangements of the integrand and the field $\xi^i$ represents one
such deformation. The connection with the exterior derivative comes through
Stokes' theorem
\begin{equation}
\int_\cM d(\xi\cdot\hat{\rho} f) = \int_{\partial \cM} \xi\cdot\hat{\rho} f
= 0,
\end{equation}
where the last equality assumes sufficient convergence at the boundary or an
empty boundary, and $\xi\cdot\hat{\rho}$ denotes the $(n-1)$-form obtained from
the interior product of the vector field $\xi$ with the $n$-form $\hat{\rho}$.
This is related to the operator $A_\xi$ through $ d(\xi\cdot\hat{\rho} f) =
\frac{1}{n}\hat{\rho} A_\xi f $.

A vector field $\xi$ was not present in the original problem of evaluating
$\esp{f}$ but its introduction is needed for a geometric formulation of the
SDE. The same phenomenon occurs in other situations, such as the problem of
finding the extrema of a function $f$ by the steepest descent. Assuming that
$f$ is a scalar, the equation $\dot x^i=-\partial_i f$ leads to different
trajectories when applied in different coordinates systems. A
coordinate-independent formulation requires instead $\dot
x^i=-g^{ij}\partial_j f$, where $g^{ij}(x)$ transforms as the contravariant
components of a metric field. The steepest descent method is closely related
to CL, so the introduction of a metric is also needed in a geometric
formulation of that stochastic process \cite{Salcedo:1993tj}. While the
previous geometric considerations have played no role in the one-dimensional
case, they are likely to be more relevant for higher dimensions.

\medskip\medskip {\bf Acknowledgments:}\\ This work has been partially
supported by the Spanish MINECO (grant No. FIS2017-85053-C2-1-P) and by the
Junta de Andalucia (grant No. FQM-225).

\newpage

\appendix

\section{Explicit counting of dimensions}
\label{appA}

\subsection{Nonperiodic densities of rational type}
\label{appA.1}

In this appendix we will study densities $\rho$ of rational type. 
Since $\rho'/\rho$ is rational, $\rho$ is holomorphic on $\C$ up to 
isolated singularities. $\rho$ can be written as
\be
\rho(z) = P(z) \exp (R(z))
\label{eq:nc_rho}
\,.
\ee
with $P$ and $Q$ rational functions. As domain we take the space of 
polynomials 
\be
\cD = \mspan \{ z^n, \, n\in \N_0 \}
\,.
\ee
For this domain, the paths connecting generalized zeroes can be naturally
grouped into equivalence classes and $\Gamma$ will include one path for each
class, plus one path encircling each isolated singularity. Certainly the weak
and strong decay conditions in Definition 4 are fulfilled by polynomials, at
finite zeroes and at infinity when the latter is an essential singularity of
$\rho$.

In (\ref{eq:nc_rho})  $P(z)$ is a general rational function 
(general up to a global factor) which we write as
\be
P(z) = \prod_{\ell=1}^{N_p} p_\ell(z)^{\alpha_\ell},
\qquad
p_\ell(z) = z-a_\ell,
\quad \alpha_\ell \in \Z \backslash \{0\}
,\quad
N_p \ge 0
\label{eq:P}
\ee
and the $a_\ell$ are pairwise different. On the other hand, $R(z)$ is also 
a general rational function, but expressed differently, for convenience:
\be
R(z) = Q(z) + R_s(z),
\label{eq:R}
\ee
where $Q(z)$ is a polynomial of degree $N_q$
\be
Q(z) = \sum_{k=0}^{N_q} c_k z^k,
\qquad
N_q \ge 0 ,\quad
c_{N_q} \not=0
,
\ee
and $R_s(z)$ contains the principal parts:
\be
R_s(z) = \sum_{m=1}^{N_s} \sum_{r=1}^{\beta_m} \frac{d_{m,r}}{q_m(z)^r}
,
\qquad
N_s \ge 0
,
\label{eq:Rs}
\ee
with
\be
q_m(z) = z-b_m,
\quad \beta_m>0,
%\quad \beta_m\in\Z,
\quad
d_{m,\beta_m} \not=0
,
\ee
and the $b_m$ are pairwise different. 

Due to invariance under translations, we can assume without loss of generality
that the $a_\ell$ and the $b_m$ are all different from zero.

In order to prove the equality
\be
N_\Gamma = N_{\mathrm{SDE}},
\ee
the first task is to compute the number $ N_\Gamma$ of independent paths 
in $\Gamma$. Independence means here that the corresponding linear 
functionals $T_\gamma$ form a maximal linear independent set. Some of the 
issues involved were illustrated with in Example 2 in Section 
\ref{sec:4}. 

Considering now the general case (\ref{eq:nc_rho}), let $N_p^\prime$ be the
number of $a_\ell$ which are different from the $b_m$, $N_p^\prime \le
N_p$. The positive $\alpha_\ell$ correspond to zeroes and the negative ones to
poles. Further, there are $\beta_m$ zeroes from each $b_m$ in $R_s(z)$ and
$N_q$ zeroes at infinity from $Q(z)$, plus $N_s$ essential singularities from
$R_s(z)$. So the total number of zeroes plus finite singularities is
\be
N_g = N_p^\prime + N_q + \sum_{m=1}^{N_s}(\beta_m+1)
.
\ee
This gives
\be
N_\Gamma = \left\{ \begin{matrix} N_g-1 &\text{$\rho$ has zeroes} \\
N_g & \text{$\rho$ has no zeroes} \\
\end{matrix}\right.
\,.
\ee
$\rho$ has no zeroes iff $N_q=N_s=0$ and all $\alpha_\ell<0$. One final remark
is that in the special case of $N_s=0$, $P(z)$, and hence $\rho(z)$, vanishes
at infinity if $\sum_\ell\alpha_\ell<0$. Such a zero is ineffective in the
domain of polynomials because $\lim_{z\to\infty} z^n\rho$ would not be zero 
for all $n$. Equivalently, only zeroes common to all functions in 
$\rho\cD$ are effective. Thus such would-be zero does not contribute to 
$N_\Gamma$.

To compute $N_{\mathrm{SDE}}$ we consider a $T\in\cH^*$ and impose the SDE
\be
0 = \esp{Az^n} = \esp{ n z^{n-1} + v(z)z^n } ,\qquad n \ge 0
\label{eq:SDE1}
\ee
where we have used the notations
\be
\esp{f} \equiv (T,f)
\ee
and
\be
v(z) \equiv \frac{\rho^\prime}{\rho} = 
\sum_{\ell=1}^{N_p} \frac{\alpha_\ell}{p_\ell}
+\sum_{k=1}^{N_q} k c_k z^{k-1}
- \sum_{m=1}^{N_s} \sum_{r=1}^{\beta_m} \frac{rd_{m,r}}{q_m^{r+1}}
.
\label{eq:v}
\ee
One can see that the coefficients of $T$ involved in the SDE (\ref{eq:SDE1})
are of the type
\be
E_n \equiv \esp{z^n},
\qquad
F_\ell  \equiv \esp{1/p_\ell(z)}
,\qquad
G_{m,r}  \equiv \esp{1/q_m(z)^r}
.
\ee
The coefficients $E_n$ are sufficient for $T$ acting on $\cD$ but $F_\ell$ and
$G_{m,r}$ are required on $\cH=\cD+A\cD$.

The expressions involved can be analyzed making use of the expansion
\be
\frac{z^n}{q(z)^r} = 
%\frac{(q+b)^n}{q^r} = 
\sum_{j=0}^n \binom{n}{j}q^{j-r} b^{n-j}
,\qquad
q(z) \equiv z-b,
~
r\ge 1, ~ n \ge 0,
~ b\not=0
,
\label{eq:binom}
\ee
or, schematically, simplifying the notation by not writing the 
coefficients of the powers of  $z$ and $q$:
\be
\frac{z^n}{q(z)^r} \sim
\left\{\begin{matrix}
\frac{1}{q^r} + \cdots + \frac{1}{q^{n-r}} & \quad n<r \\
\frac{1}{q^r} + \cdots + \frac{1}{q} + 1 + z + \cdots + z^{n-r} & \quad n \ge r 
\end{matrix}\right.
,
\qquad 
r\ge 1, \quad n \ge 0
\,.
\ee
Hence, schematically, the various terms in $\esp{Az^n}$ produce,
\begin{equation}
\begin{split}
\esp{n z^{n-1}} &= n E_{n-1} \,, \\
\esp{\alpha_\ell z^n/p_\ell} &\sim F_\ell + E_0+\cdots + E_{n-1} \,, \\
\esp{k c_k z^{n+k-1}} &= kc_kE_{n+k-1} \,, \\
\esp{rd_{m,r}z^n/q_m^{r+1}} &\sim 
\left\{ \begin{matrix}
G_{m,r+1} + \cdots + G_{m,r-n+1} & \quad n \le r \\
G_{m,r+1} + \cdots + G_{m,1} +E_0 + \cdots + E_{n-r-1}& \quad n > r \\
\end{matrix}\right. \,.
\label{eq:13.14}
\end{split}
\end{equation}
Clearly, for large enough $n$ just one new term appears when $n$ increases by
one unit, thus we definitely obtain a proper recursion from the SDE, i.e., a
finite-dimensional $\Ann A\cD$.

Momentarily we assume generic values for the parameters in $\rho$. In
particular, we assume that no $a_\ell$ coincides with any $b_m$, so
$N_p^\prime = N_p$ and all the $F_\ell$ are different from the
$G_{m,1}$. Expanding $\esp{Az^n}$ with (\ref{eq:binom}), one can see from
(\ref{eq:13.14}) that for $n=0$, the following expectation values are involved
in the recursion
\be
(n=0) \qquad
\{F_\ell \},
~
\{
E_0,\ldots, E_{N_q-1} 
\},
~
\{
G_{m,2},\ldots G_{m,\beta_m+1}
\}
\ee
The number of coefficients is $N_p+N_q +\sum_m\beta_m$, with one constraint
among them, hence $N_p+N_q +\sum_m\beta_m-1$ free parameters. For $n=1$, the
coefficients related by the recursion are
\be
(n=1) \qquad
\{F_\ell \},
~
\{
E_0,\ldots, E_{N_q} 
\},
~
\{
G_{m,1},\ldots G_{m,\beta_m+1}
\}
\ee
This introduces anew $E_{N_q}$ and all the $G_{m,1}$ ($N_s$ values), and one
constraint, hence $1+N_s-1$ new free parameters. In general for $n\ge 1$, the
formula involves
\be
(n\ge 1) \qquad
\{F_\ell \},
~
\{
E_0,\ldots, E_{N_q+n-1} 
\},
~
\{
G_{m,1},\ldots G_{m,\beta_m+1}
\}
.
\ee
For $n\ge 2$, each time a new coefficient is introduced (namely
$E_{N_q+n-1}$) which is fixed by the new constraint. The number of degrees of
freedom is thus (from $[n=0] + [n=1] + \cdots$)
\be\begin{split}
&[N_p+N_q+\sum_m\beta_m -1] + 
[ 1+N_s-1] + [1-1]+ \cdots + [1-1] +\cdots
\\&
=
N_p+N_q + \sum_{m=1}^{N_s}(\beta_m+1)-1 = N_g-1
,
\end{split}\ee
i.e.,
\be
N_{\mathrm{SDE}} = N_g-1
\ee
which coincides with the counting of $N_\Gamma$ independent paths. This
analysis covers the case of generic $\rho$ of the class rational function
times exponential of rational function.

For the non generic cases, when some of the $a_\ell$ coincide with some of the
$b_m$, hence $F_\ell=G_{m,1}$ for some pairs $(\ell,m)$, those $G_{m,1}$
appear already at $n=0$ so they do not count as new coefficients in $n=1$. This
reduces $N_p$ to $N_p^\prime$ in the counting, which again implies
$N_{\mathrm{SDE}} = N_\Gamma$.

We do not make an exhaustive analysis of all particular cases but will analyze
in more detail the cases of the type $\rho(z)=P(z)$, i.e. a rational
function. For these densities $N_q=N_s=0$ and $N_g = N^\prime_p=N_p$. The SDE
give
\be\begin{split}
(n=0) \qquad 0 &= \sum_\ell \alpha_\ell F_\ell 
, \\
(n=1) \qquad 0 &= (1+\sum_\ell \alpha_\ell ) E_0 
+ \sum_\ell \alpha_\ell a_\ell F_\ell 
,\\
(n=2) \qquad 0 &= (2+\sum_\ell \alpha_\ell ) E_1 
+ \sum_\ell \alpha_\ell a_\ell E_0
+ \sum_\ell \alpha_\ell a_\ell^2 F_\ell 
,
\\
& \cdots
\end{split}\ee
The equation for $n=0$ gives $N_p-1$ free parameters in the recursion and in
general the remaining equations do not change this. There are two cases:

1) First $\sum_\ell\alpha_\ell \ge 0$ (hence some $a_\ell$ must correspond to
zeroes instead of poles). The coefficient in front of $E_{n-1}$ in the $n$-th
equation, namely, $n+\sum_\ell\alpha_\ell$, is not vanishing and the value of
$E_{n-1}$ is determined from the recursion, for all $n$. Thus no new free
parameters are introduced and $N_{\mathrm{SDE}} =N_p-1$, which coincides with
$N_\Gamma$ in this case.

2) The other possibility is $\sum_\ell\alpha_\ell < 0$. In this case the
coefficient of $E_{n-1}$ vanishes for the equation with
$n=-\sum_\ell\alpha_\ell$. Generically that equation eliminates (fixes) one
free parameter (among the $F_\ell$ and $E_k$, $0\le k \le n-2$) but $E_{n-1}$
is a new free parameter, hence still $N_{\mathrm{SDE}} =N_p-1$. This is
correct when not all $\alpha_\ell$ are negative, therefore $\rho$ has zeroes
and $N_\Gamma=N_p-1$. However, when all $\alpha_\ell$ are negative the
equation for $n=-\sum_\ell\alpha_\ell$ turns out to be redundant (a
consequence of the previous ones)\footnote{We do not have a closed proof of
  this. It is trivial when $N_p=1$, and it is also easily proven when all
  $\alpha_\ell=-1$ (i.e., simple poles only) but it holds in all cases
  analyzed by us.}  and no free parameter is eliminated, hence
$N_{\mathrm{SDE}} =N_p$, which again matches $N_\Gamma$, since $\rho$ has only
poles and no zeroes.

\subsection{Periodic densities of rational type}

Here we consider periodic densities of rational type. The separate 
treatment is needed as some differences appear in the present case.

The class of densities considered is as follows
\be
\rho(z) = \omega^\gamma P(\omega)\exp(R(\omega))
\,,
\qquad
\omega\equiv e^{iz}
\,.
\ee
Here $\gamma\in\Z$, and $P(\omega)$ is the same function (with respect to
$\omega$) as in (\ref{eq:P}) with the added condition that the $a_\ell$ cannot
be zero (the factor $\omega^\gamma$ takes care of that). Further, $R(\omega)$
obeys (\ref{eq:R}) where $R_s(\omega)$ there is as in (\ref{eq:Rs}), with the
new condition that all the $b_m\neq 0$. On the other hand, $Q(\omega)$ is 
slightly more general:
\be
Q(\omega) = \sum_{k=N_q^-}^{N_q^+} c_k \omega^k,
\qquad
N^\pm_q\in\Z, \quad
N_q^- \le N_q^+
,\qquad
c_{N^\pm_q} \not=0
.
\ee

The domain will be the space spanned by the Fourier modes in $[0,2\pi]$
\be
\cD = \mspan \{ e^{inz}\,|\, n\in\Z \} = \mspan \{ \omega^n\,|\, n\in\Z \}
\,,
\ee
and $\Gamma$ contains one path of each equivalence class.

$N_\Gamma$ is the total number of closed paths plus the number of zeroes minus
one, unless there are no zeroes. This number can be computed as follows. The
(finite) singularities come from $R_s(\omega)$ and $P(\omega)$. This gives a
number of closed paths
\be
N_c = N_s + \sideset{}{'}
\sum_{\ell=1}^{N_p} \Theta(\alpha_l<0) + 1
\,,
\ee
where the function $\Theta(x)$ takes the value $1$ (resp. $0$) when the
proposition $x$ is true (resp. false) and the prime indicates to exclude the
term if $a_\ell$ equals some $b_m$ in $R_s(\omega)$. The plus one counts the
closed path encircling once the cylinder.

The number of finite zeroes is
\be
N_{z,\mathrm{finite}}  = 
\sum_{m=1}^{N_s} \beta_m + \sideset{}{'}
\sum_{\ell=1}^{N_p} \Theta(\alpha_l>0)
\,.
\ee
Let us count the number of zeroes of $\rho$ at $z=-i\infty$ (i.e.,
$\omega=\infty$).  The factor $Q(\omega)$ gives $(N_q^+)_+$ zeroes, where
$(x)_+\equiv \max(x,0)$. The factor $\omega^\gamma P(\omega)$ would give one zero
when $\gamma+\sum_l \alpha_l<0$ which could be effective only if $(N_q^+)_+=0$
(otherwise the essential singularity dominates). However, such zero from
$\omega^\gamma P(\omega)$ does not contribute even when $(N_q^+)_+=0$ because
$\omega^n \rho$ would not go to zero for sufficiently large $n$. Thus
\be
N_{z,-i\infty}  =  (N_q^+)_+ \,.
\ee
Similarly for $z=+i\infty$ ($\omega=0$)
\be
N_{z,+i\infty}  =  (-N_q^-)_+
\,.
\ee
Collecting the various contributions
\be
N_\Gamma = (N_q^+)_+ + (-N_q^-)_+ + N_p^\prime + \sum_{m=1}^{N_s}(\beta_m+1)
 +1 - \Theta(N_z>0)
\,,
\label{eq:NGamma}
\ee
where $N_z=N_{z,\mathrm{finite}}+N_{z,-i\infty}+N_{z,+i\infty}$ is the total
number of zeroes. As before in the noncompact setting, $N_p^\prime$ denotes
the number of $a_\ell$ in $P(\omega)$ different from any $b_m$ in
$R_s(\omega)$. It is noteworthy that our treatment is asymmetric, after
choosing to express $\rho(z)$ in terms of $\omega=e^{iz}$ instead of
$e^{-iz}$, yet $N_\Gamma$ is unchanged under $z\to-z$, as it should.

In order to compute $N_{\mathrm{SDE}}$, let us apply $A$ to a generic element
of the basis,
\be
-iA\omega^n = 
(n+\gamma)\omega^n + \sum_{\ell=1}^{N_p} \alpha_\ell\frac{\omega^{n+1}}{p_\ell}
+\sum_{k=N_q^-}^{N_q^+} kc_k\omega^{n+k}
-\sum_{m=1}^{N_s}\sum_{r=1}^{\beta_m}rd_{m,r}\frac{\omega^{n+1}}{q_m^{r+1}}
\,.
\label{eq:Aomega}
\ee
For $T\in\Ann A\cD$, the SDE imply $\forall n\in\Z ~
\esp{-iA\omega^n}=0$. As in the noncompact case, the action of $T$ is fully
determined in terms of the basic coefficients 
\be
E_n \equiv \esp{\omega^n},
\qquad
F_\ell  \equiv \esp{1/p_\ell(\omega)}
,\qquad
G_{m,r}  \equiv \esp{1/q_m(\omega)^r}
.
\ee
To do the reduction, for positive $n$ the expansion in (\ref{eq:binom})
directly applies (with $\omega$ instead of $z$) while for negative $n$ the
following expansion applies
\be\begin{split}
\frac{\omega^n}{q(\omega)^r} &= 
\sum_{j=1}^{-n} \binom{-r}{-n-j}\frac{(-b)^{n-r+j}}{\omega^j}
+
\sum_{j=1}^{r} \binom{n}{r-j}\frac{b^{n-r+j}}{q^j}
,\\%\qquad
& q(\omega) \equiv \omega-b,
~ 
r\ge 1, ~ n < 0, ~ b\not=0
.
\label{eq:binomneg}
\end{split}\ee
Using these expansions, from inspection of (\ref{eq:Aomega}) one obtains the
following general structure
\be
\{E_n\}, \{E_{n+N_q^-},\ldots, E_{n+N_q^+} \},
\ee
from the first and third terms in the r.h.s. of (\ref{eq:Aomega}). The second
term gives
\be\begin{split}
n\ge 0 &\qquad \{ F_\ell, E_0,\ldots,E_n \},  \\
n=-1 &\qquad \{ F_\ell \},  \\
n\le -2 &\qquad \{ F_\ell, E_{n+1},\ldots,E_{-1} \} 
\,,
\end{split}\ee
and the fourth term gives ($1\le r \le \beta_m$)
\be\begin{split}
n\ge r &\qquad \{ G_{m,r+1},\ldots,G_{m,1},E_0,\ldots,E_{n-r} \},  \\
-1\le n<r &\qquad \{  G_{m,r+1},\ldots,G_{m,r-n}\},  \\
n\le -2 &\qquad \{ G_{m,r+1},\ldots,G_{m,1},E_{n+1},\ldots,E_{-1} \} 
\,.
\end{split}\ee
Again, as in the noncompact case, for large $|n|$ the number of new terms
increases at the same rate as $|n|$ hence this is a proper recursion, with
$N_{\mathrm{SDE}}<\infty$.

Rather than analyzing all cases, let us consider a density $\rho$ with
parameters taking generic values, plus the explicit simplifying assumption

$N_q^-<0<N_q^+$. For $n=0$ the coefficients involved are
\be
(n=0) \qquad
\{F_\ell,E_{N_q^-},\ldots, E_{N_q^+}, G_{m,1},\ldots,G_{m,\beta_m+1}\}
\,,
\ee
therefore the number of free parameters is
\be
N_p + (N_q^+-N_q^-+1) + \sum_{m=1}^{N_s}(\beta_m+1) -1
\,.
\ee
For $n=1,2,\ldots$ or $n=-1,-2,\ldots$ 
one finds that each time just one new coefficient is introduced, which gets
determined by the equation. Thus no new parameters are generated and
\be
N_{\mathrm{SDE}}=  N_p + N_q^+-N_q^- + N_s + \sum_{m=1}^{N_s}\beta_m
\,.
\ee
As expected, this number matches $N_\Gamma$ in (\ref{eq:NGamma}).

The examples 1 and 3 analyzed in Section \ref{sec:3} illustrate non 
generic cases not covered by this discussion.

\section{Vanishing of the net flux through the boundary of a region in a 
steady state}
\label{appB}

Let $P(z)$ be the stationary state of the CL process for a  
density $\rho(z)$. $P$ is normalized and we assume that $P$ obeys the 
Fokker-Planck equation (\ref{eq:fp}); if  $v$ has singularities, 
(\ref{eq:fp}) has to be understood as a weak solution (see below). 
(\ref{eq:fp}) is just the continuity  equation of the CL process: 
\begin{equation} 
0 = \partial_t P(z;t) + \partial_x j_x + 
\partial_y j_y 
\end{equation} 
with density current 
\begin{equation} 
j_x = 
v_xP - \partial_x P, \qquad j_y = v_y P\, 
. 
\end{equation} 
By the divergence theorem for any closed curve $\gamma$ that is the 
boundary of a region $G$ free of singularities of $v$, $\gamma=\partial G$ 
and avoids any singularities of $v$, the net flux of $j$ through $\gamma$ 
vanishes.

The same holds in the case of a periodic $\rho$ for any closed loop 
$\gamma$ homotopic to $z(t)=t+iy_0$ ($y_0$ a constant value and 
$t\in[0,2\pi]$) and avoiding any singularities of $v$. In this case the 
manifold is the cylinder $[0,2\pi]\times \R$ and $\gamma$ divides the 
cylinder in two disconnected regions $C_+=\{(x,y), y>y_0\}$ and 
$C_-=\{(x,y), y<y_0\}$ with $\gamma$ as common boundary.  Each region 
carries a fraction of the normalization of $P$, denoted $\alpha_\pm$ 
respectively, with $0 \le \alpha_\pm \le 1$ and $\alpha_++\alpha_-=1$. The 
CL process is a driven Brownian process, hence its sample paths can be 
chosen to be almost everywhere continuous. Therefore the CL walkers can 
only enter or leave the regions $C_\pm$ by crossing the boundary $\gamma$. 
If the net flux through $\gamma$ were not zero, the fractions $\alpha_\pm$ 
would change under time evolution, which is not possible for a stationary 
state.  As a consequence, $\int_0^{2\pi} dx j_y(x,y_0)=0$ and likewise 
for any path homotopic to the one written.

If the path $\gamma$ encloses a singularity of $v$, we need the assumption 
that $P$ is a weak solution of the Fokker-Planck equation, which makes the 
vanishing of the net flux through $\gamma$ obvious. To justify the 
assumption, consider a small closed path enclosing a singularity; based on 
experience from CL simulations (see for instance \cite{Aarts:2017vrv}), 
we expect $P$ to vanish at least linearly at the singularity; this implies 
that the current $j$ remains bounded there and $P$ is indeed a weak 
solution of (\ref{eq:fp}). Letting the enclosing path shrink to zero then 
implies that there is no net flux coming in or out of singularities.

The argument can be extended to conservative processes in any number of 
dimensions to state that the flux through a closed hypersurface 
(codimension 1) must vanish in a stationary state. The conditions to 
obtain such a statement are i) the stochastic process must have almost 
everywhere continuous sample paths, ii) the stationary density must be 
normalizable,
% {\bf isn't that guaranteed, as the total probability is  always 1?}
and iii) the hypersurface must divide the manifold in two regions.

Obviously continuity is needed, even to have a well-defined density
current. The normalizability condition is also relevant. Consider for instance
the cylinder of the periodic case as the manifold and a constant density there
with a constant density current in the $y$ direction. The flux through the
line $y=0$ is not zero even if the system is conservative. The statement does
not apply because the fractions $\alpha_\pm$ are undefined when $P$ is not
normalizable. Finally, the condition that the hypersurface divides the
manifold is also relevant. Consider for instance a two-dimensional torus as
manifold, from compactification of $[0,2\pi]\times [-Y,Y]$. Again one can have
a constant density and a constant flux in the $y$ direction without violating
flux conservation, yet the flux through the line $[0,2\pi]$ does not
vanish. In this case the statement does not apply because the line $[0,2\pi]$
does not separate the torus into two regions and the quantities $\alpha_\pm$
are meaningless.

\end{document}